\documentclass[sigconf]{acmart}
\AtBeginDocument{%
  \providecommand\BibTeX{{%
    \normalfont B\kern-0.5em{\scshape i\kern-0.25em b}\kern-0.8em\TeX}}}

\copyrightyear{2021}
\acmYear{2021}
\setcopyright{rightsretained}
\acmConference[FAccT '21]{Conference on Fairness, Accountability, and
Transparency}{March 3--10, 2021}{Virtual Event, Canada}
\acmBooktitle{Conference on Fairness, Accountability, and Transparency
(FAccT '21), March 3--10, 2021, Virtual Event,
Canada}\acmDOI{10.1145/3442188.3445872}
\acmISBN{978-1-4503-8309-7/21/03}

\begin{document}

\title{Representativeness in Statistics, Politics, and Machine Learning
}


\author{Kyla Chasalow}
 \affiliation{
   \institution{Cornell University}}
 \email{kec89@cornell.edu}

\author{Karen Levy}
\affiliation{
\institution{Cornell University}}
  \email{karen.levy@cornell.edu}

%%\author{Anonymous 1}
%%\affiliation{%
%%  \institution{Anonymous University, Department of Anonymous}}
%%\email{anonymous2@anon.edu}

%%\author{Anonymous 2}
%%\affiliation{%
%%  \institution{Anonymous University, Department of Anonymous}}
%%\email{anonymous2@anon.edu}

\renewcommand{\shortauthors}{Chasalow and Levy}

\begin{abstract}
  Representativeness is a foundational yet slippery concept. Though familiar at first blush, it lacks a single precise meaning. Instead, meanings range from typical or characteristic, to a proportionate match between sample and population, to a more general sense of accuracy, generalizability, coverage, or inclusiveness. Moreover, the concept has long been contested. In statistics, debates about the merits and methods of selecting a representative sample date back to the late 19\textsuperscript{th} century; in politics, debates about the value of likeness as a logic of political representation are older still. Today, as the concept crops up in the study of fairness and accountability in machine learning, we need to carefully consider the term's meanings in order to communicate clearly and account for their normative implications. In this paper, we ask what representativeness means, how it is mobilized socially, and what values and ideals it communicates or confronts. We trace the concept’s history in statistics and discuss normative tensions concerning its relationship to likeness, exclusion, authority, and aspiration. We draw on these analyses to think through how representativeness is used in FAccT debates, with emphasis on data, shift, participation, and power.

\end{abstract}




\maketitle

\section{Introduction}                               
In the study of fairness, accountability, and transparency in machine learning, there have been various attempts to catalogue and unpack different understandings of the terms ``fairness'' and ``bias'' \cite{Mulligan,narayanan2018translation,Suresh,mitchell2018prediction}. Less attention has been afforded to ``representative,'' a foundational yet ambiguous word peppered through discussions of data and statistics. Newspapers allow ``nationally representative'' surveys to speak for the state of public opinion and society. A machine learning paper speaks of selecting ``representative examples'' to further interpretability \cite{ribeiro2016should}. Another warns of costs to ``representative and consultative'' participation when ML technologies scale across contexts \cite{Sloane}. Meanwhile, a legal scholar warns against relying on statistical models developed on ``unrepresentative'' reference data to make sentencing decisions \cite{Hamilton}. Another describes systematic ``underrepresentation'' in large datasets used to allocate goods and services as a rising form of civic and political exclusion \cite{lerman2013big}.

Linguistic precision is key to meaningful debate, yet some words at the heart of contemporary discussions of data, algorithms, and their consequences resist a single precise definition. Words like ``fair,'' ``biased,'' ``interpretable,'' and ``representative'' are valuable yet challenging for their multiple meanings and ability to capture intuitive concerns about technical topics. Lipton and Steinhardt warn against the overuse in machine learning of ``suitcase words'' (a term coined by Minsky) which ``pack together a variety of meanings'' and have colloquial appeal \cite{LiptonStein, Minsky}. While suitcase words can be useful for overarching ideas, they require careful unpacking to make it clear when we are talking past one another or eliding distinctions that have significant conceptual and normative consequences. In this paper, we unpack representativeness.

Fundamentally, representativeness concerns the ability of one thing to stand for another---a sample for a population, an instance for a category. We could think of all of statistics along these lines: as a web of philosophies and practices for describing and making inferences about the world through inevitably limited data. However, unlike many terms of statistical inference---multicollinearity, maximum likelihood, posterior means---``representative'' and its complement, ``unrepresentative,'' appear intuitive and familiar. They engage people in thinking about the trustworthiness and quality of data and the inferences based on them. 

In fact, representativeness has long been a contested concept in both statistics and politics. In statistics, debates about the merits and methods of selecting a representative sample date back at least to the late 19\textsuperscript{th} century. As Kruskal and Mosteller have traced, the term has been persistently slippery \cite{KandM1, KandM2, KandM3, KandM4, Kruskal78}. Its uses within statistics have not been siloed from its more colloquial use to mean typicality, a miniature, the ``absence of selective forces,'' ``general acclaim'' for data, or some notion of coverage \cite{KandM1}. In politics, the adjective and noun ``representative'' is older still, developing in the 17\textsuperscript{th} century the sense of  ``standing for others'' that has become central to representative government \cite{Williams}. There is an evident connection to representation, another tricky word notably explored by Pitkin in \emph{The Concept of Representation} (1967) \cite{Pitkin}. Pitkin locates ``representativeness'' in the logic of descriptive likeness. This notion of political representation has some affinity with the statistical sample, a connection made explicit in random selection of jury pools, some forms of deliberative ``minipublics,'' and occasional calls for random sampling of government representatives \cite{Callenbach1985, Pitkin, Fung2003}. It also, however, raises a persistent debate about the value of likeness relative to other logics of political legitimacy \cite{Pitkin, Manin}.

Since the 1970s, psychologists have also studied representativeness as part of how we think and judge. Tenenbaum and Griffiths call it a ``central explanatory construct'' used to explain ``phenomena of categorization, comparison, and inference,'' including ``errors in probabilistic reasoning'' \cite{Tenenbaum2001}. The latter refers to the representativeness heuristic, introduced by Tversky and Kahneman to describe our tendency to make judgements based on \emph{similarity} in ways that violate the rules of probability \cite{Tversky1981, Tversky1974}. Other work has focused on understanding our tendency to view some objects as more typical or as better examples of a category than others \cite{Mervis1981, Abbott2011}. But while that work examines representativeness as a cognitive process, representativeness is also an analytic criterion and demand in statistics and politics. We focus on that criterion here. 

In this paper, we explore uses of ``representative'' in the past and present to make sense of its meanings and their relevance for fairness, transparency, and accountability in machine learning. In the words of C.S. Lewis, we look to meet the word ``alive'' \cite[p. 2]{Lewis}. We ask not only what representativeness means but how it is mobilized socially and what values and ideals it communicates or confronts. Our goal is not to pick a best meaning or propose a technical formulation nor to pose representativeness as a panacea. Still, we must be careful when we speak of representativeness, as different meanings have distinct practical and normative implications.

We proceed as follows. In section 2, we trace the history of representative sampling, from early debates about partial data to the emergence of probability sampling. In section 3, we broaden our scope to other scientific contexts, where the word’s historical link to probability sampling provokes debate about different logics of generalizability. In section 4, we argue that representativeness raises normative tensions concerning likeness, exclusion, authority, and aspiration. We use historical examples to illustrate these concepts before pivoting to the present. In section 5, we consider representativeness in FAccT as it connects to data, shift, participation, and power.\footnote{We acknowledge our own representativeness concerns. We contribute a sense of the variety of meanings and values associated with representativeness, yet do not claim to analyze an exhaustive or representative set of them. We focus more on social sciences than natural sciences and on Europe and the United States more than other parts of the world. Beyond these scope restrictions, issues related to our selection of sources are harder to pinpoint. We take some refuge in being illustrative, yet this does not fully resolve the issue, for what we choose to include---or not---will always shape the reader’s sense of the subject.}

We see this effort as a complement to other work in FAccT that draws on fields like STS to interrogate taken-for-granted terms and categories \cite{Hanna,Hu2020}. The notion of representativeness presupposes classification and equivalence---that is, the ability for some units to be treated as interchangeable with others, at least in a given context for some salient characteristic \cite{Desr1991}. Those equivalences are essential to our ability to learn about the world, but they are also approximate, value-laden, and at times, contested \cite{Bowker, Camargo2009}. Only by considering the ways we translate the world into data in the first place can we fully examine the realities data represent and potentially, perpetuate.

\section{The Statistical Roots of Representativeness}                 
The roots of representative sampling lie in inductive inference. As Stephan wrote in a 1948 history, ``All empirical knowledge is, in a fundamental sense, derived from incomplete or imperfect observation and is, therefore, a sampling experience'' \cite{Stephan1948}. But while the act of drawing conclusions from partial information and assessing their validity is as old as the human condition, formal probability-based sampling methods were novel at the turn of the 20\textsuperscript{th} century. 

That is not to say that pre-20\textsuperscript{th} century quantitative work was devoid of partial data and inference. Lacking census data, 17\textsuperscript{th} and 18\textsuperscript{th} century political arithmeticians extrapolated from local records and their own surveys for insurance and policy purposes \cite{Stephan1948, Porter1986}, using, for example, data on local birth rates to estimate total population size \cite{Desr1998}. These forays into inference from partial data were controversial and remained so into the 19\textsuperscript{th} century, when the growth of statistical bureaucracies and regular census-taking in many nations shifted the emphasis of the emerging field of statistics from partial surveys to complete enumeration \cite{Desr1998, Gigerenzer1989}. 

Nineteenth century statistics also featured a tension between notions of statistics as a descriptive science and as a science for discovering laws and causes. While adherents of the former emphasized ``one-to-one correspondence between observations and numbers,'' with ``numbers…interpreted as summaries of immediately observable facts,'' those working in the latter framework began to calculate aggregates and make inferences and predictions \cite[p. 47-51]{Schweber2006}. Though not discussed as such at the time, we might view these tensions as early debates about the possibility for representativeness---for a move beyond that one-to-one correspondence. They also provide context for why, when representative sampling was first formally proposed to the International Statistical Institute (ISI) by statistician Anders Nicolai Kiaer in 1895, some viewed the prospect as ``dangerous,'' with one critic opining that ``one cannot replace by calculation the real observation of facts'' \cite{KandM4}.

\subsection{Representativeness as Design Problem and Sociopolitical Problem} 
By the turn of the 20\textsuperscript{th} century, representative sampling began to be conceptualized as a more formal design problem---an issue to be approached via ``systematically designed plans'' for data collection \cite[p. 37]{Kiaer1897}. Kiaer presented his experiments with sampling as evidence that it was possible to obtain an ``approximate miniature'' of the population, capturing not just ``average conditions'' but ``information on variation and extreme values'' in a rigorous way \cite[p. 53]{Kiaer1897}. Historians and statisticians have distinguished the subsequent development of sampling from earlier uses of partial data by its formal procedure and its application of probability theory \cite{Stephan1948, Seng1951}. Yet part of what was new in this time was also the speed, type, and detail of information that sampling began to provide. ``By shrinking drastically the delay between request and delivery of information,'' Prévost and Beaud write, sampling ``allowed for the production of data that would have had no meaning in the old time frame'' \cite[p. 154]{Prevost2012}. Together with the growth of statistical bureaucracies and new mechanical tabulation techniques for census-type data, these data expanded the ``knowledge-gathering capacities of modern states.''

However, the representative method itself also emerged from changing economic and political demands on statistical ``descriptions of the world'' and related changes in thinking about parts and wholes \cite[p. 211]{Desr1998}. While earlier monography work intended to illustrate ``holistic'' truths by selecting the average type, sampling emerged from a need for information on \textit{variation} among individuals \cite{Desr1998, Mespoulet2002}. As local charity solutions to poverty became inadequate in the context of national economic crises, surveys needed to provide not holistic knowledge of the typical family but nationally applicable knowledge of the condition of individuals so that governments could allocate and distribute aid. In the American context, the shock of the Great Depression ``ruptured'' established links between ``statistics and political action'' and created a need for a ``new numerical reading of America'' that could support a more interventionist government \cite[p. 4, p. 81]{Didier2020}. Businesses and election campaigns with expanding reach also demanded national-level information at low cost and high speed. Thus, as Desrosi{\`e}res argues, debates about how to obtain a representative sample of a population could only be formulated once “the problem itself…the constraint of representativeness” became salient \cite[p. 210-1]{Desr1998}. Representativeness was not only a design problem; it was also a sociopolitical problem translated into new ways of doing statistics.

\subsection{Achieving Representativeness: Mechanical and Purposive Selection}
By the 1920s, initial debate among statisticians about the possibility of sampling had largely given way to the question of \emph{how} to sample \cite{Bellhouse1988}. Given support for both random and purposive sampling methods, a 1926 ISI report endorsed both \cite{Jensen1926}. Then, in 1934, a landmark paper by Jerzy Neyman helped establish random over purposive methods, formalized stratified sampling, and contributed to the theoretical foundation of probability sampling \cite{Bellhouse1988, KandM4}. The full story of these ideas and their adoption is more layered, involving interactions among theory, practical experimentation, political demands, and bureaucratic processes \cite{Didier2020}; for our purposes, we highlight a few main theoretical tensions about how to achieve---and what it means to be---a representative sample.

``It will be obvious,'' Kiaer wrote in 1897, ``that the representative method can be applied in several ways'' \cite[p. 43]{Kiaer1897}. He described two samples, which, though not labeled ``mechanical'' or ``purposive,'' foreshadowed these logics. The first sample is premised on arbitrary filtering, including by the first letter of surnames. He reasoned that such selection should occur in a ``haphazard or random way'' so as ``to avoid, in the most stringent manner, any procedure that could give preference to persons in certain occupations or belonging to particular social strata'' \cite[p. 39]{Kiaer1897}.  This mention of randomness and preference suggests a notion of mechanical objectivity \cite{Daston1992,porter1996trust}, in the sense that by following rules in a ``stringent manner,'' representativeness would result from the absence of selective discretion. 

The same appeals underlie random sampling, a bedrock of probability theory. While Kiaer does not mention probability, statisticians would soon begin to emphasize its role and the value of estimating ``probable limits of error'' \cite{Bowley1906, Caradog1921}. The ``first essential'' of sampling, wrote Bowley in 1910, is ``that every member of the group considered should have nearly the same chance of being included in the sample'' \cite[p. 56]{Bowley1910}. It was standard early on to emphasize this ``equal chance'' as a defining feature of mechanical approaches \cite{Bellhouse1988}, perhaps because it suggests a form of procedural fairness (some even spoke of ``fair samples''  \cite{KandM4, Mcnemar1940, Bowley1915, Stephan1941}) and neutrality. After all, random samples are only guaranteed to be accurate and non-preferential on average. As Stephan reflected in 1939, ``It provides but little aid and comfort to the victim of a poor draw in random sampling to assure him that ‘in the long run’ the random method of selection will give him errors in one direction just as often as it will the others'' \cite{Stephan1939}. The issue of how to check a \emph{particular} sample for representativeness remained a point of debate \cite{KandM4, Jensen1926, Mcnemar1940, Garrett1942}. But locating representativeness in a mechanical \emph{procedure} did provide a way to move past the specter of chance skew and uncheckable variables. In the words of statistician Margaret Hogg, a firm proponent, ``only if a random sample has been secured can one argue from representativeness which has been tested, to representativeness where no test is possible'' \cite{Hogg1930}.

The ``victim of a poor draw'' in random sampling might favor another approach: using available data to ensure a proportionate match on known relevant variables. In Kiaer’s second sample, surveyors in rural areas used census data to allocate counts per county and then selected districts within counties to ``represent the main industry-groups within the country as well as its various geographic conditions'' \cite[p. 41]{Kiaer1897}. On the ground, enumerators were trusted ``to select respondents in a representative manner ... [and] check that not only typical middle-class houses were visited but also those occupied by the more well-to-do and the poorer classes'' \cite[p. 42]{Kiaer1897}. Later, surveyors collected additional data to correct non-matching proportions on variables such as occupation---a safeguard against enumerator bias. It seems to be the match that made the sample ``representative,'' not the exact selection rule or persons selected. 

What about variables for which no matching is possible? Kiaer provides only a general assurance that samples will yield ``the same degree of accuracy when it comes to new fields.'' He follows with an analogy:  ``The relationship between these two sets of returns can be illustrated by the relationship between two barrels, one small and the other very large… If a number of samples taken from each, crosswise and lengthwise in every conceivable direction, show the composition of the content of both barrels to be practically the same, I think it is justified, on the basis of further, more extensive analysis that we assume can only be carried out on the smaller barrel, to draw valid inferences about the unknown content of the larger barrel'' \cite[p. 51]{Kiaer1897}. This barrel analogy suggests what we call the \emph{universal representativeness ideal}---a proportionate match for every variable at once. Its ambition is to scale down without loss of information. Its allure lies not only in a correct answer to an immediate question. Better yet, it promises the possibility of data reuse via a kind of representativeness that holds from every angle.\footnote{Analytically, this ideal is impossible. Not only can we not check every possible variable, but universal representativeness would also have to apply to joint distributions of variables. There is also an infinite regress issue: if the whole contains a miniature, why should not the miniature contain a miniature? Yet at some point, the reduced version would become so small that we would not soon call it representative \cite{KandM1}.}

Purposive sampling formalized the idea that a sample matching on some variables would match on others; the added ingredient was correlation. Formally, it would come to involve matching to average values of known variables (``controls'') thought to be correlated to unknown variables of interest \cite{Jensen1926, Jensen1928}. What especially set it apart from mechanical approaches was its ability to incorporate \emph{existing knowledge}. To repeat a pedagogic example discussed in \cite{Neyman1934}, if we wished to learn the average right arm length of a group of people and we knew the lengths of their left arms, we might select people whose left arms were closest to the average left arm length. The fact that left and right arm lengths are highly correlated would likely lead to a fairly accurate result. In applied problems, correlations are often smaller than left-right arm correlations are, and practitioners who attempted the method encountered problems \cite{KandM4, Neyman1934, Coats1931}. It was stratification that would become a foundational method for incorporating existing knowledge into probability sampling. 

\subsection{Stratification: A Representative Method}
Although stratification appeared in some earlier sampling work \cite{Mespoulet2002, Bowley1926}, Neyman helped establish it for a wider audience \cite{Bellhouse1988}. Stratification bridged the purposive-random tension, providing a way to increase the precision of estimates by incorporating knowledge of group differences into random sampling \cite{Desr1998, Stephan1941}. It was also part of the emerging theory of probability sampling, in which unequal chances of selection were allowed and sometimes preferable. Today, a probability sampling design refers to a specification of possible samples and their probabilities of selection. Equal probabilities of inclusion are not required; what matters is that the probabilities of inclusion of each unit are known so that they can be used as weights \cite{Lohr1999}. The literature that has built on these ideas over the last eighty or so years is expansive, but for our purposes, the interesting question is how these ideas relate to representativeness. 

A less discussed feature of Neyman’s work is the way that he relocates representativeness to a property of a method. ``Obviously,'' he wrote in 1934, ``the problem of the representative method is \emph{par excellence} the problem of statistical estimation,'' later adding, ``If there are difficulties in defining the ‘generally representative sample,’ I think it is possible to define what should be termed a \emph{representative method of sampling} and a \emph{consistent method of estimation}'' \cite[p. 572]{Neyman1934}. By attaching ``representative'' to ``method'' and not to ``sample,'' Neyman avoids the uncertainties inherent to any one sample and focuses on building a general theoretical framework for inference that performs well on average and allows estimates of precision. In the process, he broadens the scope of ``representativeness'' to apply to any legitimate probability-based design---a sample could be disproportionate but result from a representative method. 

    In practice, people still speak of representative \emph{samples}, but when applied to probability sampling in general, it seems that accurate maps from sample to population are the basic ingredient for representativeness \cite{KandM2, KandM3}. It then makes sense to speak of a sample being ``weighted to be representative'' (e.g. \cite{pew}), perhaps restoring a sense of proportionality. If, by design or implementation, some subgroups of the target population lack any representatives, if weights are inaccurate, or if other distortions arise that impede inference, then the sample (or method) might be called unrepresentative (e.g. \cite{Christodoulou2020, Aronow2016}). Admittedly, this is broad, and it remains an open question how sample size, degree of uncertainty, and other methodological elements should factor in here. But it is also flexible. In a way, we have come full circle, back to the basic link between statistical representativeness and scientific inference.

\section{Representativeness in Science}                                
Most broadly, representativeness is a generalizability concern. But generalizability is itself a suitcase word, packing in notions of sample-to-population inference, transfer across populations, and external validity from experiments to real-world settings \cite{LiptonStein, salganik2019bit, Elwood2013, Firestone1993}. Within that assortment, some use representativeness primarily to describe sample-to-population inference yielding time- and place-specific descriptions of population frequencies, as in much survey research. Here, the question of whether a sample is representative of a target population may be complemented by a scope of inference question: what are the data at hand actually representative of \cite{KandM3}? In contrast, the question of whether, for example, biological findings in fruit flies apply to humans or of whether the results of experiments with college students apply to other people introduce generalizability logics different from the probability sampling one. It is important to distinguish these logics, but two realities complicate those distinctions.

First, in practice, representativeness is sometimes used more flexibly to describe patterns of exclusion and inclusion across objects of scientific study. In clinical trials, for example, the word is used to describe the over- and under-representation of minorities and other consistent selection biases, which raise both generalizability and equity concerns \cite{Britton1999, Fisher2011}. Similarly, in genetic research, the fact that people of European ancestry make up a large portion of data used to study disease genetics has consequences both for underrepresented groups and for science \cite{Goldberg2020, Sirugo2019}. These examples concern (un)representativeness across fields of research and their collections of studies, datasets, and subjects. A related flavor of the concept is used to discuss the ``model organism'' problem. The term refers to the organisms commonly used in biology, such as \emph{Drosophila melanogaster} (the fruit fly) \cite{Bolker1995}. It has been extended as a metaphor for when a field relies heavily on certain research objects (Twitter in social media research \cite{tufekci2014big}, chess in AI \cite{Ensmenger2012}), with consequences for the field’s direction and findings. Though often simple and accessible, model organisms and collections of model organisms may, in part because of these traits, exclude certain kinds of organisms relative to the ‘population’ of possible objects of study. The possibility for such exclusions to shape the outcomes and causal pathways we observe broadens the scope of representativeness, suggesting it can underlie other generalizability logics.

The second complication is that distinctions between types of generalizability are more than semantics---they are sites of methodological contestation. Thus the choice of whether and how to use ``representative'' is not purely a matter of its technical meaning. For example, a recent debate among epidemiologists about the importance of representativeness in their field turns out to also be a debate about the scope and focus of that field \cite{Rothman2013a, Ebrahim2013, Nohr2013, Elwood2013, Richiardi2013a}. ``Exalted along with motherhood, apple pie and statistical significance,'' one author writes, ``representativeness may be essential for conducting opinion polls, or for public-health applications, but it is not a reasonable aim for a scientific study,'' adding that the former are ``not science in the same way that causal studies about how nature operates are science'' \cite{Rothman2013a}. Another paper pushes back, allowing representativeness a somewhat broader scope beyond random sampling but also defending the scientific value of ``descriptive epidemiology'' \cite{Ebrahim2013}. In qualitative research, debate about the use of ``representative'' to describe case selection touches on long-running tensions around the infringement of quantitative logics and standards \cite{becker1996epistemology, small2009many}. ``Representative'' is a bridging word---occurring in many fields and providing a common language for specialist and non-specialist concerns---but it is also used to draw boundaries.

The preceding discussion creates ambiguities concerning how far the word ``representative'' extends or should extend. On one extreme, it has a fairly specific historical link to probability sampling. On the other, representativeness is flexible, arising in discussions of over-, under-, non-, or mis-representation of descriptive characteristics or causal mechanisms, for particular samples and for collections of many kinds. This variety suggests a need for caution. There is value in the term’s flexibility, but this paper is not a call to pepper our prose with the word---quite the opposite. If anything, researching representativeness has convinced us that we should be sparing in its use, or at least, that ``representative'' can rarely stand alone. 

\section{Normative Tensions of Representativeness}            
(Un)representativeness has a fascinating capacity to describe scientific uncertainties, statistical methods, principles of government, and social (in)equities---sometimes together. So far, we have focused on representativeness mainly as a matter of accurate inference and scientific concerns. We now pivot to a more sociopolitical and normative perspective, for representativeness is not only a descriptive fact; it is also a value, ideal, and mark of authority---merited or not. Here, we untangle some of the normative tensions that surround the concept, illustrating them with historical examples, before turning in section 5 to consider representativeness in the FAccT context. 

\subsection{The Likeness Ideal}

Within her taxonomy of representation, Pitkin locates representativeness in descriptive resemblance---in standing ``for others ‘by being sufficiently like them’''\cite[p. 80]{Pitkin}. As we have seen, applied to data, this suggests a proportionate match or perhaps just a known relationship between data and some target of inference. Applied to social and political processes, it suggests a value for likeness between people and their representatives on relevant variables. It is this that gives concerns about the representativeness of these processes some affinity with statistical concepts of representativeness. 

Political processes often do not involve everyone. Representativeness promises to reconcile those exclusions with principles of equity \cite[p. 130]{Manin}. For example, the fact that not all people are activists may be more acceptable if activists are deemed ``representative'' of those who do not participate---thereby preserving all individuals’ right to be heard \cite{Verba}. This promise of restoring a sense of inclusion reflects two logics---one normative and one epistemic \cite{ONeill2001}. First, representativeness may be a normative end linked to voice and democratic legitimacy, as well as to the idea that those who share relevant identities or experiences with the people they represent are better advocates for those people---what Phillips termed the ``politics of presence'' \cite{Phillips1995}. Even when it does not translate to direct action, likeness may have symbolic value, inspiring trust \cite{Miller2020}. Second, representativeness may be an epistemic means to inclusive and generalizable knowledge about people’s views and experiences. These epistemic goals may still be motivated by equity issues and, since knowledge may be tied to identity and experience, are sometimes served by the ``presence'' logic  \cite{harding1992rethinking}.

While the appeals described above may be more prominent in political contexts, they can also apply to data, particularly when data are a form of sociopolitical representation. In either case, the value of likeness may be contested. In political representation, it confronts other logics of political legitimacy, including those that emphasize ``knowledge, expertise, or judgement'' or authorization and accountability more than shared identity \cite{ONeill2001}. Here, representativeness as likeness may be irrelevant or rejected in favor of other goals. As described in section 3, the value of representative data may also be contested. But as we now illustrate with a case from the history of sampling, even when some form of representativeness \textit{is} the goal, it can be complicated by other values and logics.

In 1862, concerned about farmers’ vulnerability to speculators spreading false information, the \emph{American Agriculturist} published a call for ``readers….in every town [to] counsel together, and select some man who may be relied upon for good judgement, and general ability to estimate with some degree of accuracy in regard to the leading crops'' so that together, a few hundred or more reports from across the country could give ``an approximation to the average of the whole'' \cite{Judd1862}. This method was soon adopted by the new U.S. Department of Agriculture, which cultivated long-term relationships with a network of volunteer crop reporters valued for their intellect and public spirit \cite{Didier2020}. These volunteers were not intended to be \emph{like} other farmers---indeed, they were selected precisely because they were deemed to be \emph{unlike} other farmers in their exceptional discernment of their and their neighbors' crop yields. Accurate data were to result from a kind of epistemic coverage.

However, in the 1920s, problems with the method’s accuracy---attributed in part to reporters being misled by their neighbors \cite{Didier2020}---led to the ``individual-farm'' approach, in which farmers reported only on their own farms \cite{Harlan1939}. This transformed farmers from \emph{spokespeople} to \emph{examples}, making it necessary to consider their representativeness \cite{Didier2020}. Here, the same exceptional status that had legitimized agricultural data before became an obstacle. The eventual solution was probability sampling. The cost was a new problem of nonresponse, for ``respondents could refuse to answer, but could not refuse to be selected'' \cite[p. 195]{Didier2020}. Data’s role as a form of political representation changed, too. Because random selection precluded sustained relationships and because, in the New Deal context, data were meant more for planning government interventions than for farmers’ use, the move engendered more separation between data collector (government) and data subject (farmer). In a sense, ``representativity had replaced participation'' \cite[p. 209]{Didier2020}.

The case of the crop reporters illustrates a number of tensions. While crop reporting reflects the democratic appeal of participatory data collection, it is also exclusionary, reflecting a political logic that emphasizes knowledge and skill over likeness. In contrast, designed sampling reflects the appeal of purposeful, structured inclusion, but at a cost to voluntary participation. These tensions between design and volunteerism and between representatives selected to be exceptional or broadly inclusive will not be new to social and political scientists---but they manifest anew, for example, in debates about crowdsourcing today \cite{salganik2019bit, stamm2017, aitamurto2014self}. Finally, the case highlights that the inclusiveness of a representative sample is no guarantee of power over how data are used. Broadly, then, representativeness, rooted in the idea of one thing standing for another by ``being like'' it, interacts with other values and logics. It is already difficult to define and achieve representativeness, yet it is also crucial to ask what it achieves and whether it is the right goal in the first place.

\subsection{Exclusionary Representativeness}
 
Recall the universal representativeness ideal suggested by Kiaer’s barrel analogy: a perfect miniature on every variable. Jorge Luis Borges identifies the impossibility of this ideal in his short story ``The Congress'' (1977), in which an idealistic group attempts to devise a Congress to represent all people everywhere. The group quickly encounters intractable obstacles in deciding which variables matter: ``Alejandro Glencoe might represent not only cattlemen but also Uruguayans, and also humanity’s great forerunners, and also men with red beards, and also those who are seated in armchairs. Nora Erfjord was Norwegian. Would she represent secretaries, Norwegian womanhood, or---more obviously---all beautiful women? Would a single engineer be enough to represent all engineers---including those of New Zealand?'' \cite{borges1977book}. The variables multiply, and it turns out that the only perfect ``Congress of the World'' is the world itself.  In practice, we must decide what variables are salient to evaluate for their representativeness. Epistemic and normative goals can lead to different criteria here---the former concerned with whether variables correlate with those of interest, the latter concerned with social or political relevance \cite{ONeill2001}. It is worth emphasizing that in either case, these goals can be shaped by our ideals---and that, despite the link between representativeness and inclusion, such ideals can have exclusionary consequences.

Consider the case of \emph{Middletown}, a 1929 study of Muncie, Indiana by Helen and Robert Lynd, notable at the time for applying an anthropological lens to everyday American life \cite{Lynds, Igo2007Avg}. The over-five-hundred-page book was mostly qualitative, interspersed with snippets of interviews with residents on ``getting a living,'' ``making a home,'' ``using leisure,'' and more, but throughout, occasional tabulated data helped reinforce its status as an objective empirical study. The book was an unexpected bestseller. Though widely received as an unprecedented window into a representative American community, the Lynds were cautious but inconsistent regarding \emph{Middletown}’s representativeness \cite{Igo2007Avg}. They warned that ``a typical city, strictly speaking, does not exist,'' but did write that one of their selection criteria was ``that the city be as representative as possible of contemporary American life'' \cite[p. 3]{Lynds}. Of course, they also called the study ``Middletown,'' a name which abstracts it from Muncie and suggests the ordinary and average, ideas susceptible to being transmuted into typical and representative.

\begin{figure}[h]
  \centering
  \includegraphics[width=7cm]{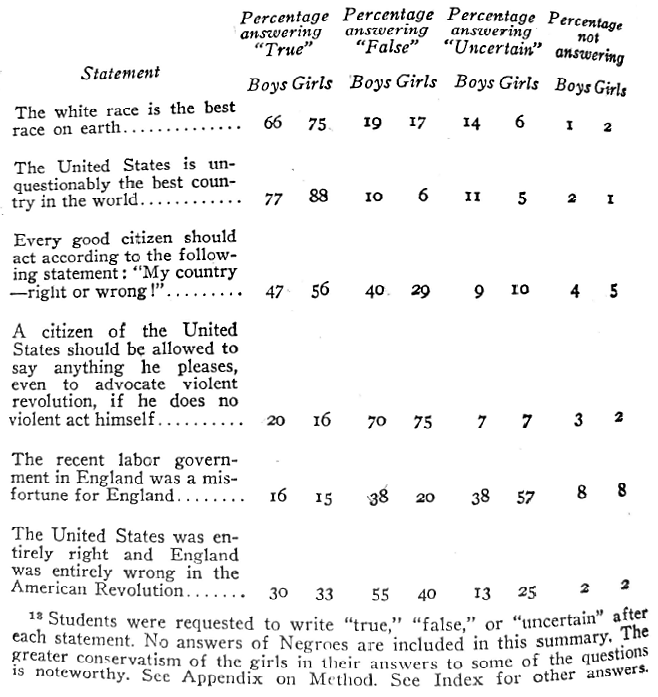}
  \caption{Though \emph{Middletown} was praised for its objectivity, it was more than a recording of facts. An especially stark example is the inclusion of a question asking schoolchildren about their endorsement of white supremacy while ``excluding the answers of Negroes'' \cite[p. 200]{Lynds}.
 }
 
  %\Description{?}
  \label{middletownImg}
\end{figure}

As Igo has traced, the Lynds then make a puzzling acknowledgment: Muncie is ``unusual'' \cite{Igo2007Avg}. They say they have selected for a city of moderate size with “a small Negro and foreign-born population,” adding that to avoid dealing at once with racial and cultural change, ``it seemed a distinct advantage to deal with a homogeneous, native-born population, even though such a population is unusual in an American industrial city'' \cite[p. 8]{Lynds}. They see no contradiction. Muncie is both ``as representative as possible'' and unusual. In fact, there was a growing black population in Muncie, but the Lynds excluded them from their results (see e.g. fig. \ref{middletownImg}) \cite[p. 56-7]{Igo2007Avg}. We might label \emph{Middletown} unrepresentative of Muncie or America, but we should also ask in what way the Lynds could suggest it \emph{was} representative. The answer, Igo writes, is that ``the Lynds’ representative community was less an empirical than a normative proposition'' \cite{Igo2007Avg}. Their goals \emph{were} epistemic but their concept of their object of study was shaped by an ideal of a ``purer, simpler, even preindustrial, America,'' especially the white, Protestant part of it---an ideal more ``wished-for than real'' \cite[p. 59]{Igo2007Avg}.

Admittedly, Middletown is different from the statistical samples we discussed earlier. It is a single case study raising a ``representativeness as typicality'' logic and ironically, therefore particularly susceptible to being in some way atypical.\footnote{``Typical'' is yet another common but ambiguous word. It is sometimes used synonymously with representativeness but has more sense of being characteristic, ideal, or both---concepts that are exclusionary in that they deemphasize variety \cite{Kruskal78, Daston1992}.} But Igo finds a similar tendency in early Gallup opinion polls, which she suggests under-sampled African Americans, immigrants, women, and people with lower incomes in part because they did not fit pollsters’ normative concept of what the likely voter \emph{should be} (reality did come back to bite in the 1948 election) \cite[p. 138]{Igo2007Avg}. These cases illustrate a form of representativeness claim defined relative to ideals and prescriptions rather than empirical data alone. This does not render the empirical data useless, but those data do exist within a framework in which some variables and individuals are deemed more relevant than others to what it means to be representative. These considerations are always present to some degree as we confront the immense ``Congress of the World.''

\subsection{Rhetorical Authority}
A third tension surrounding the concept of being ``representative'' concerns the authority that the word itself can inspire, merited or not. The Lynds \emph{were} somewhat cautious about making representativeness claims. The public reception to \emph{Middletown} was less so. Those who did not read it themselves were liable to encounter it via the newspaper, church, or community event, where its findings were received as if they reflected the state of America \cite{Republican1929, Montclair1929, Chatham1929, Miami1929a, Miami1929b}. The \emph{St. Cloud Times} of Minnesota opined that: ``The picture of ‘Middletown’ is representative of millions of our population, groping for a national culture. The social surveys are unbiased and scientific. It is a book, which should be read by all thoughtful Americans'' \cite{CloudTimes1930}. Such claims seem to be less about specific numbers or descriptions and more about a vague sense that \emph{Middletown} captured ``us'' (or at least, those of ``us'' who fit its ideals) \cite{Igo2007Avg}. This public reaction reflects the way representativeness can take on a life of its own, capturing the imagination while also deriving authority from its apparent status as an objective scientific property. This poses a challenge for communication. Words can travel beyond their original intent. Mundane and resonant, imprecise yet somewhat technical, ``representative'' may be especially liable to do so.

\begin{figure}[h]
  \centering
  \includegraphics[width=3.5cm]{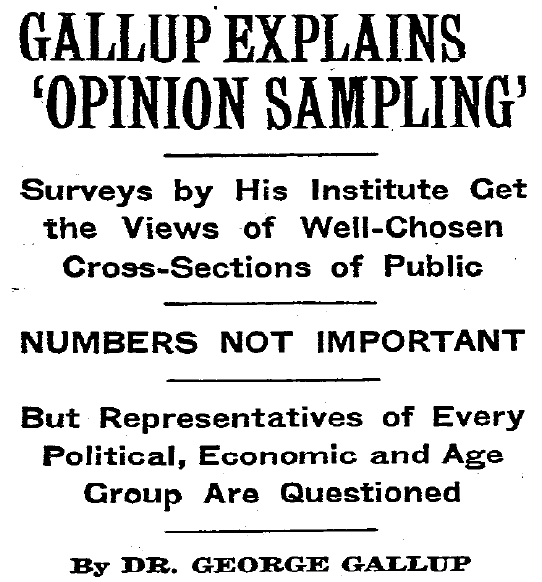}
  \caption{ The headline of a 1938 \emph{New York Times} article \cite{Gallup1938NYT}.}
  %\Description{?}
  \label{gallupImg}
\end{figure}

Moreover, the word’s authority can be wielded strategically. We see this, for example, in the way George Gallup advertised his new opinion polls to the American public in the 1930s and 40s. Gallup did more than present statistical theory. He translated statistical representativeness into political representation, writing of miniature electorates, sampling referenda, and representatives chosen from every group (see fig. \ref{gallupImg}) and framing the polls as a way to support democracy by restoring the ``voice of the common man'' \cite[p. 13]{GallupPulse}. At the same time, he emphasized the polls’ scientific legitimacy to distinguish them from the public failure of the \emph{Literary Digest} polls in the 1936 election, a failure he attributed to a misplaced focus on sample size rather than representativeness \cite{GallupPulse}. ``New polls… based on scientific principles'' prioritized the ``cross section,'' enacting the principle that ``the most important requirement of any sample is that it be as representative as possible of the entire group or ‘universe’ from which it was taken'' \cite[p. 57]{GallupPulse}.\footnote{Though Gallup spoke of random sampling here, in early years, he used quota sampling, which set target proportions to mirror important divisions in the population of interest and then charged interviewers with finding people to fill the targets.}

    The problem with the rhetorical authority of the word ``representative'' is that it can be unmerited or misleading. It is misleading to present results without addressing their representativeness, but it is also misleading to suggest representativeness without specifying in what way and to what extent (even then, those cautions are sometimes ignored). The word’s rhetorical halo can also mask important dynamics, including, in the case of Gallup polling, the question of who the data in question most benefited. Gallup’s enthusiastic rhetoric of the democratic ``sampling referendum'' does not account for the reality that polling had close links to market research and served the needs of businesses, media, and political parties---at times more ``surveillance'' than ``voice'' \cite{Beniger1983, Edwards2011}.  

\subsection{Time and Aspiration}

Data are inherently historical. Barring time travel, it is impossible to collect data on the future. As Nohr and Olsen have written, “Representativeness is time- and place-specific, and will therefore always be a historical concept. Representativeness is gone as we speak, as Heraclitus told us more than 2000 years ago: ‘You cannot step twice into the same river’” \cite{Nohr2013}. In this spirit, we turn now to questions about representativeness and time.

Bouk describes a late 19\textsuperscript{th} century struggle between African Americans and major American life insurance companies that raises temporal and normative concerns which implicate representativeness \cite{Bouk2015}. In the 1880s, in the wake of the Civil War and Reconstruction period, thousands of African Americans submitted applications for life insurance. At first, insurance companies were mostly willing to sell. The catch: African Americans were charged the same rate as white applicants but for only two-thirds of the standard benefit. This, the insurers argued, was not a race-based decision but a statistical one. The companies drew on the 1870 census and other data to show that African Americans had higher mortality rates, an unsurprising finding given recent slavery and ongoing discrimination. The insurers also went further, arguing that these differences applied not only to the current time but always. Emphasizing their reliance on the facts, the life insurers draped racial prejudice with a statistical veil.  Bouk introduces the term “fatalizing” to describe how insurers used past data to assign African Americans to a statistically backed fate.

Beginning in 1884, activists succeeded in pushing for anti-discrim\-ination laws in a number of Northern states. Their arguments had two bases. First, they opposed fatalizing practices by defending the possibility for mortality rates to change. “Where life insurers initially assumed continuities between the slave-era past and a free future,” Bouk writes, “African Americans and their allies interpreted the Civil War as a moment of rupture, severing the past from the present and future” \cite[p. 32-3]{Bouk2015}. The question was not so much whether the data used by life insurers were accurate at the time they were collected as whether those data could stand for the present and future. Second, however, advocates emphasized equality in insurance as “a matter of right” \cite[p. 41]{Bouk2015}. Even if mortality rates were different at present or if they remained so at some future point, the principle of equality trumped even accurate fatalism.

This case helps us consider the time-dependence of representativeness in three ways. First, consider the link between the past and the present. The case raises an \emph{empirical} concern about change: even data representative in the past may not be representative in the present if the object they describe has changed. But second, what if past data \emph{do} accurately capture the present, but that present still bears the mark of historical discrimination and injustice? Here we confront an \emph{aspirational} tension concerning future possibility and what some term “historical bias”---“a normative concern with the world as it is” (and how it might be different) that applies even when we sample and measure perfectly \cite{Suresh}. Insurers’ fatalizing claims not only obscured any change that might have already occurred; they also worked against change by reinforcing existing inequities. In prediction especially, where the object is inherently a future event, feedback loops and future possibility raise questions about what it means for past data to, in a sense, ‘stand for’ the future, together with a normative objection to being represented even by data that may, at present, be representative.

Finally, when advocates argued that non-discrimination in insurance was a “matter of right,” they implied that rights and aspirations can be more important than data---representative or not. This echoes our earlier assertion: that representativeness confronts other values, logics, and goals and that the concept’s intuitive link to inclusion is not straightforward. Overall, we emphasize that the time-dependence of representativeness is an empirical and normative matter. The decision to accept past data as representative can be a matter of accuracy---how much has the present situation shifted relative to the past?---but also of aspiration---do the data reflect the future we would like to see?

\section{Representativeness in FA\lowercase{cc}T}
Recall that formal sampling emerged in the early 20\textsuperscript{th} century not only as a design question posed by statisticians but also as a sociopolitical demand for new kinds of data that people translated into new ways of doing statistics \cite{Desr1998}. Similar dynamics emerge today. In machine learning and the FAccT community, we again see social problems translated into new ways of working with data, with the potential both to exacerbate social problems and to understand and address them \cite{abebe2020roles,kleinberg2018discrimination}. Within that space, representativeness is not a new concern, but its relationship to FAccT concepts and values should be carefully considered. Drawing on what we have learned from tracing the concept through statistical history, scientific discourse, and sociopolitical perspectives, we map four areas where the concept arises in FAccT---data, shift, participation, and power.

\subsection{Data}

At the turn of the 20\textsuperscript{th} century, mechanical tabulation and sampling helped expand government and commercial data collection, creating an influx of data---so much so that in 1905, Mandello worried that the “ever increasing” demand for statistics had yielded “a mass of uncontrollable figures for the uncontrolled use of everybody” \cite{Mandello1905}. Today, we again live amid an influx, and it has become common to speak of “big data,” characterized by new scale, speed, formats, and computational approaches. While some have hailed a move away from “small data” probability sampling \cite{Mayer2013}, others have defended its importance \cite{Couper2013, Morstatter2014} or emphasized that representativeness-related questions of selection bias, missingness, and generalizability remain crucial for big data \cite{boyd2012critical, Olhede2018}. In a similar vein, though the “GIGO principle” (garbage in; garbage out) is well-known, some have argued that we need to do more to evaluate and communicate where datasets used in ML come from \cite{selbst2019fairness,gebru2018datasheets,denton2020bringing, paullada2020data}. 

In machine learning, representativeness raises new flavors of ambiguities we encountered earlier. These include the question of whether “representative data” refers to a proportionate match to some target population or to a more flexible sense of being able to learn “generalizable predictive patterns” \cite{Bzdok2018}. The idea of learning predictive patterns lies at the heart of ML, but to link these patterns to the words “representative” or “generalizable” is a nontrivial step that can obscure distinctions. Two examples illustrate the danger. 

First, data summarization algorithms are sometimes described as reducing data to a “representative set” \cite{Borovicka2012, Celis2018, Pan2005}. In an echo of the earlier split between mechanical and purposive (judgement-based) selection, they do not \emph{per se} sample randomly but rather seek to intentionally remove redundancies while preserving relevant information---in effect, to isolate the signal from the noise. Crucially, however, ``representative” here concerns a subset of the original dataset and not necessarily that larger dataset relative to some external target---something the word “representative” by itself does not tell us. Second, the goal of distinguishing meaningful patterns from noise also motivates the practice of evaluating models on holdout data to detect overfitting to training data---often described in terms of measuring ``generalization error” and improving ``generalizability” to other data from the same distribution \cite{LiptonStein, Borovicka2012}.\footnote{When we split into train and test sets, ``we expect that a representative sample will be in each subset'' \cite{Borovicka2012}, showing how embedded the idea of random selection as representative has become. But “representative” here is relative to the original dataset.}  Yet we should distinguish that a non-overfit model, while less sensitive to variations in the data, may still capture an unrepresentative (or non-generalizable) overall pattern relative to some \emph{intended} whole or deployment context. There is a risk here that “generalizable” or “representative” appear to promise more than they do---all the more so given that the external target or deployment context is exactly what people often use ``representative'' to mean, as a kind of synonym for ``real-world" testing and applicability (e.g. \cite{albert2020, paullada2020data}).

In FAccT, further ambiguities ensue when representativeness mixes with suitcase words such as “fairness” and “bias” and the related phrases “fair representation”  and “representation bias.”  “Representation bias” can refer to at least two distinct issues: (1) a disproportionate sample relative to a true distribution; or (2) a class imbalance problem, in which smaller groups may face worse model performance even when sampled proportionately \cite{Suresh}. This latter concern is one common notion of unfairness \cite{chouldechova2018frontiers}, and it suggests that if representativeness means a proportionate match, it might at times yield \emph{unfair} representation. Yet some do use “representative” to discuss equal representation. To that end, Buolamwini and Raji coin the phrase “user-representative sets” to describe datasets that have “equal representation of each distinct subgroup of the user population”  (which might involve oversampling smaller groups) \cite{raji2019actionable}. Such data are “representative,” perhaps, in that they ideally contain sufficient data to allow accurate prediction for members of all relevant groups---a kind of coverage. Of course, even this form of representativeness is not the cure for all forms of bias or unfairness, which also depend on modeling decisions and sociopolitical context \cite{GebruTutorial,isbell2020keynote,raji2019actionable}. Data, while essential and sometimes overlooked, are only one piece of that puzzle.

So far, we have focused on the single dataset, but what was true of clinical trials and model organisms is also true here: representativeness may concern \emph{collections} of research objects. Machine learning and FAccT research itself often rely on various commonly used datasets, which act as benchmarks for comparing models and methods \cite{hand2006,buolamwini2018gender, Young2019}. If benchmarks do not “represent the target population,” this can create “evaluation bias,” rewarding models that perform well on unrepresentative data \cite{Suresh}. More fundamentally, these datasets form what Denton \emph{et al.} call an “infrastructure” for the field, acting “analogously to model organisms” and becoming “stand-in[s] for more complicated data traces and machine learning tasks” \cite{denton2020bringing}. In the process, they may individually or collectively reproduce selection tendencies as well as classification and measurement decisions across studies and applications (e.g. \cite{shankar2017classification}). 

\subsection{Shift}
  
A basic assumption in supervised machine learning is that data distributions match across training, testing, and deployment. As Hand writes, “Intrinsic to the classical supervised classification paradigm is the assumption…that future points to be classified are drawn from the same distributions as the design set,” adding pessimistically that “the assumption that the design distribution is representative of the distribution from which future points will be drawn is perhaps more often incorrect than correct” \cite{hand2006}. Even if a design distribution \emph{is} initially representative, as noted earlier, representativeness is time- and place- dependent. This fact becomes all the more important in ML given (1) the tendency for ML-driven systems to be applied across contexts, driven in part by the financial advantages of scalability \cite{Sloane} and (2) the fact that these systems are implemented in dynamic environments where, “in more extreme cases, actions influenced by a model may alter the environment, invalidating future predictions” \cite{lipton2018mythos}. These tendencies are also part of how we critique these systems, warning of “portability trap[s]” \cite{selbst2019fairness} and context-dependence \cite{Sloane}, of “zombie predictions” unable to account for policy change \cite{koepke2018danger}, or of “lack of representativeness” as a source of injustice when risk tools developed on one population are used to make decisions about another \cite{Hamilton}. 

While not all portability challenges over time and space can be detected in terms of “shifts in the joint distribution of features and labels” \cite{selbst2019fairness}, a growing body of work seeks to address such deviations from the “representativeness” or “matching distribution” assumption in machine learning. When only a few “out of distribution” data points are at play, practitioners speak more of outlier or anomaly detection, but when those points reflect meaningful changes in the underlying data-generating distribution, the problem is conceptualized as “shift” \cite{Storkey2009,rabanser2019failing,shafaei2018does,koh2020wilds}. The goals of this work are to detect and model various forms of shift for the sake of adjustment or at least “failing loudly” when a model receives inputs that do not come from the distribution on which it was trained \cite{rabanser2019failing}. Though translating that theoretical work to actively detecting and responding to shift in commercial, bureaucratic, and policy contexts remains a challenge, shift work \emph{is} motivated by an empirical question: do distributions match? 

Yet as we saw in African Americans’ struggles against discrimination in life insurance, the prospect of \emph{the future} throws another wrench in this picture. Of course, if we are to predict at all, we must assume some link between data and the unobservable future, but that link raises normative dilemmas. Again, the world may not be as we wish it to be and people may push for desirable and aspirational outcomes not yet reflected in data \cite{kleinberg2018discrimination, Kay2015, Celis2018}. “We” is a bit deceptive here: people may disagree on what those outcomes should be or on the morality of predictive practices \cite{kiviat2019moral}. These disagreements have high stakes, for their outcomes are performative---that is, acting on data today (or not), even when they are currently, by some standard, accurate and representative, shapes future possibilities.

\subsection{Participation}

Representativeness finds further expression in calls for participatory processes intended to ensure equitable and inclusive ML-driven decision-making. A growing area of work urges greater opportunity for stakeholders---especially members of affected communities---to shape ML systems \cite{Sloane,lee2019webuildai,martin2020participatory}. For example, the call for papers for the 2020 ICML Workshop on Participatory Approaches to Machine Learning decried the fact that “the design choices of a few” often determine how a system operates and espoused “more democratic, cooperative, and participatory” processes that “encourage the perspectives of those impacted by an ML system” \cite{participatory}. Lee \emph{et al.} design a  framework to “promot[e] representative participation” in order to “empower stakeholders who typically do not have a say in the algorithms that govern their services, communities, or organizations” \cite{lee2019webuildai}. Sloane \emph{et al.} describe “representative and consultative participation” as a means of obtaining the “right information from the right people,” though they argue such participation is not a panacea for ML’s ills \cite{Sloane}. Here, we again see representativeness linked to both political and epistemic aims. 

One such aim, as we discussed in section 4.1, is to ensure the right to have a meaningful voice in designing systems that affect one’s community. Representativeness in this sense is a matter of design justice, as exemplified in the phrase “nothing about us without us” \cite{costanza2020design}. Representative processes are further espoused as an epistemic means to more effective and equitable systems, based on the prospect that community stakeholders hold substantial information about how systems are likely to work in practice and what impacts they will have on affected communities \cite{Aitamurto2016}. Without adequate representation of this knowledge in deliberative processes about ML systems, these impacts are likely to be underappreciated. Representativeness may also lend democratic legitimacy to ML systems that are tailored to community goals, rather than simply imposed “out of the box” without regard for local conditions \cite{lee2019webuildai}.

Even if we decide that a decision-making process \emph{should} be representative, we must still determine \emph{of whom} it should be representative. Defining the population relative to which we assess representativeness is a crucial decision but not an obvious one. When we espouse representative or participatory processes in ML decision-making, we typically are not aiming for \emph{national} representativeness; rather, we often have in mind some sort of \emph{targeted} representativeness of an “affected population”---that is, “populations who have direct experience with the public system in question” \cite{brown2019toward}---or a collection of “stakeholders,” a term which is itself contested \cite{mitchell1997toward}. A concomitant concern is determining what attributes are important to represent---as Borges’ allegory illustrates, this requires judgment and line-drawing. Moreover, as in data collection, even when there is a demand for participants to be “representative” of some population, it is not always clear whether this means simply the inclusion of a variety of perspectives, representation \emph{proportionate} to some population, or the \emph{over}-representation of historically marginalized views in order to amplify them \cite{diverse2017}.\footnote{What’s more, these processes have an active and ongoing quality. Proponents of participatory ML speak of having voices “at the table” \emph{throughout} decision-making processes---including system formulation, deployment, and post-hoc evaluation. We do not only “do representativeness” at a moment in time and consider the matter settled.} 

Representativeness, for all its virtues, might not achieve other normative goals. Many of the most fundamental concerns raised by algorithmic processes involve key civil and constitutional rights, like equal protection, freedom from discrimination, and due process \cite{richardson2019litigating}. Likeness-based matching---the idea that decision-makers should share attributes with the population, so as to form a “miniature” of the community (see sec. 4.1)---might be an ineffective prophylactic against the denial of these rights. Specialized expertise necessary to ensure these rights may come from institutional affiliates (like legal advocates) who work \emph{with} affected groups but are not necessarily themselves members of it \cite{diverse2017}. In this sense, some participatory processes echo the logic of early crop reporting---in which reporters were not intended to match the farmer population, but were valued as representatives for their specialized knowledge. This is not to say that expert advocates alone provide adequate representation. As Costanza-Chock demonstrates, strategies in which designers use techniques to “stand in for” engagement with the community itself risk embedding invalid assumptions about what its members want \cite{costanza2020design}. Participatory ML must consider how to balance representativeness alongside other logics and values.

Finally, participatory ML raises the key question of whether representation in a decision-making process translates to a system that benefits and meaningfully distributes power to those represented. To the degree that representativeness bestows legitimacy on an ML decision-making process, the term may be marshaled for rhetorical purposes---much as we saw in section 4.3. Just as proclaiming Gallup polls “representative” could paper over methodological details and questions about who most benefited from them, ML “participation-washing” might confer legitimacy on a decision-making process even when that process tokenizes, extracts value from, or further disenfranchises the people ostensibly represented. Debates about representativeness might also distract attention from whether the technology should be deployed at all \cite{Sloane,costanza2020design,hoffmann2020terms}.

\subsection{Power}  
We close on the issue of data and power. Representativeness is a concept and value situated in social power dynamics. Despite the positive connotation it gains from its links to representation and inclusion, the apparent inclusiveness of representative data is not a guarantee of ethical data collection or control by data subjects over how data are used, nor does it ensure that those represented in a dataset will benefit from its use. Faced with a problem of unrepresentativeness, collecting more data can be an appealing solution for governments and companies that derive power, knowledge, and value from data collection---power not only over those they directly observe in their data, but also, via statistical inference, over the populations represented by those data \cite{kiviat2019moral,barocas2020privacy}. Such data collection may benefit that population, but the pursuit of representative data may also lead to exploitation and surveillance---particularly of those least able to resist \cite{fussellattempt,hoffmann2020terms,powles2018seductive,hawkins2018beijing}. In fact, a narrow focus on collecting more data to improve a given technology could overshadow the issue of whether that data collection or the technology itself is desirable in the first place \cite{abebe2020roles,haggerty2009methodology}. The question is: representativeness for whom? For whose benefit or harm?

 The potential for data to be used for harm (even or especially when they are accurate) has driven some to call not for inclusion, better coverage, or representativeness, but for obfuscation and non-collection \cite{bedoya2014big,nabil2017,brunton2015obfuscation}---or as a 2019 workshop on these issues aptly framed it, “Please Don’t Include Us” \cite{pleasedont}. The comparative harms of inclusion and exclusion from datasets are complex, contested and contextual. We do not aim to resolve them here, and they are thoughtfully considered by other scholars \cite{benjamin2019race,hoffmann2020terms}; but for our purposes, it is crucial to note how these tensions complicate the positive connotations often associated with representativeness.

For all the difficulties of making ethical judgements about inclusion and exclusion \emph{within} the datasets we collect, questions of power also attend those we do not collect. This is, in a sense, the converse of the “model organism” datasets we discussed in 5.1; while some datasets are used widely across domains, others are not collected at all. Mimi Onuoha’s “Library of Missing Datasets” \cite{onuoha} and the Centre for Public Numbers’ “Missing Numbers” project \cite{missingnumbers} highlight the omission of certain datasets (e.g. statistics on police brutality) from otherwise data-rich environments. As Onuoha notes, missing datasets are not accidental, but may occur because powerful actors are not inclined to collect them. In response, efforts to collect \emph{counterdata} \cite{dignazio2020data} seek to fill these gaps and document phenomena that are important to marginalized groups. Sometimes, this involves not collecting more data about \emph{those} groups, but “studying up,” a notion originating in anthropology \cite{nader1972up} and recently raised in FAccT \cite{barabas2020studying}. The call to ``study up'' implores us to reorient our research questions---and, concomitantly, our datasets---by making the powerful the targets of analysis. These efforts are oriented toward a sort of field-level representativeness, aiming to ensure that the field neither underscrutinizes the powerful nor underrepresents the interests of the less powerful.

\section{Conclusion}
In exploring representativeness, our challenge has been to open the concept's ambiguities in a useful way without collapsing them to a single definition. We find that representativeness, with its roots in statistical sampling, is often a vaguely technical description linked alternatively to the typical example, random sample, proportionate miniature, and to the basic aims of scientific inference. As a political concept, it is used to communicate value for likeness and inclusion---but at times to exclude. Its combined scientific status and political resonance give it rhetorical authority, merited or not. For both data and politics, it is complicated by the roles power and aspiration play in shaping our futures. FAccT brings many of these dynamics together. But in exploring representativeness, we have also teetered at the edge of a rabbit hole, relying on additional suitcase words like fairness, normativity, and power. Hazardous as this may be, we could not do otherwise. The web of concepts that surround the concept of representativeness complicates it---but also gives it its resonance. Again: representativeness can rarely stand alone.

\begin{acks}
We thank Kim Weeden, Marty Wells, Malte Ziewitz, and Emmanuel Didier, as well as members of Cornell's AI, Policy, and Practice Initiative and College Scholar Program, the John D. and Catherine T. MacArthur Foundation, and the Miami Foundation.
\end{acks}

\bibliographystyle{ACM-Reference-Format}
\bibliography{representativeness}

\end{document}